\documentstyle[emulateapj,psfig]{article}

\makeatletter

\newenvironment{inlinefigure}{%
\def\@captype{figure}%
\noindent\begin{minipage}{0.999\linewidth}\begin{center}}
{\end{center}\end{minipage}\smallskip}
\makeatother

\lefthead{Wehner, Barger, \& Kneib}

\begin{document}
\title{The Submillimeter Properties of Extremely Red Objects}
\author{E.~H.~Wehner,$\!$\altaffilmark{1}
A.~J.~Barger,$\!$\altaffilmark{1,2,3}
J.-P.~Kneib$\!$\altaffilmark{4}
}
\altaffiltext{1}{Department of Astronomy, University of Wisconsin-Madison,
475 North Charter Street, Madison, WI 53706}
\altaffiltext{2}{Department of Physics and Astronomy,
University of Hawaii, 2505 Correa Road, Honolulu, HI 96822}
\altaffiltext{3}{Institute for Astronomy, University of Hawaii,
2680 Woodlawn Drive, Honolulu, Hawaii 96822}
\altaffiltext{4}{Observatoire Midi-Pyr{\'e}n{\'e}es, 14 Avenue 
E.~Belin, 31400, Toulouse, France}

\slugcomment{Accepted by the Astrophysical Journal Letters}

\begin{abstract}

We use deep near-infrared and submillimeter observations of
three massive lensing cluster fields, A370, A851, and A2390, to 
determine the average submillimeter properties of a $K'$-selected 
sample. The 38 Extremely Red Objects (EROs; $I-K'>4$) 
with $K'<21.25$ have a very significant error-weighted mean 
850~$\mu$m flux of $1.58\pm0.13$~mJy. The ERO contribution to 
the 850~$\mu$m background is
$1.88\pm 0.16\times 10^4$~mJy~deg$^{-2}$, or about half the 
background light. The 17 Very Red Objects 
(VROs; $3.5<I-K'<4$) are also significantly detected
($1.32\pm 0.19$ mJy), bringing the combined VRO and ERO
contribution to $2.59\pm 0.19 \times 10^4$~mJy~deg$^{-2}$.
There is a substantial systematic uncertainty (about a 
factor of two) in this value due to field-to-field variation,
but even with this uncertainty it is clear that a large fraction 
of the 850~$\mu$m background arises 
from red objects. An analysis of the VRO and ERO number counts shows 
that half of the population's 850~$\mu$m light arises in objects 
with demagnified magnitudes $K'<20$ and half in fainter objects. 
On the basis of the $I-J$ versus $J-K'$ colors of 
the galaxies, the bulk of the submillimeter signal appears to arise 
from the dusty starburst galaxies in the red object population rather 
than from the high-redshift elliptical galaxies.

\end{abstract}


\section{Introduction}
\label{secintro}

Extremely Red Objects (EROs) have been the subject of 
many studies since their discovery (e.g., 
\markcite{cowie90}Cowie et al.\ 1990; 
\markcite{hu94}Hu \& Ridgway 1994). Two scenarios can explain 
the unusally red colors of the EROs: old stellar 
populations at $z>1$ (e.g., 
\markcite{dunlop96}Dunlop et al.\ 1996;
\markcite{spinrad97}Spinrad et al.\ 1997;
\markcite{soifer99}Soifer et al.\ 1999;
\markcite{cimatti02}Cimatti et al.\ 2002)
or dust-enshrouded galaxies (e.g., 
\markcite{graham96}Graham \& Dey 1996;
\markcite{cimatti98}Cimatti et al.\ 1998;
\markcite{smith01}Smith et al.\ 2001). 
The starlight from dusty galaxies is
reprocessed and reemitted in the rest wavelength
far-infrared band. For high redshift sources, this
light is redshifted into the submillimeter. Recently,
a number of possible ERO counterparts to submillimeter detected sources  
have been discovered
(e.g., \markcite{smail99}Smail et al.\ 1999, 2002;
\markcite{barger00}Barger, Cowie, \& Richards 2000;
\markcite{ivison02}Ivison et al.\ 2002).

The {\it COBE} satellite
found that the submillimeter extragalactic background light (EBL)
has approximately the same integrated energy density as the optical
EBL (\markcite{puget96}Puget et al.\ 1996; 
\markcite{fixsen98}Fixsen et al.\ 1998); thus, a large
fraction of starlight is reradiated by dust. To determine
the high-redshift star formation, the sources that comprise the 
submillimeter EBL need to be determined.
Here we use deep submillimeter and near-infrared 
(NIR) imaging of three massive lensing cluster fields to 
determine the submillimeter properties of the ERO population.
Two advantages of observing cluster fields 
are that background sources are magnified and mean 
source separations on the sky are increased.

In \S~\ref{secdata} we briefly describe our $I$, $J$, $K'$, and 
850~$\mu$m data. The $15''$ beam size of SCUBA 
(\markcite{holland99}Holland et al.\ 1999) on 
the 15~m James Clerk Maxwell Telescope\footnote{The James Clerk Maxwell 
Telescope is operated by the Joint Astronomy Centre on behalf 
of the UK Particle Physics and Astronomy Research Council, the
Netherlands Organization for Scientific Research, and the
Canadian National Research Council} prohibits us from 
making individual matches between submillimeter and optical/NIR sources. 
However, in \S~\ref{secebl} we 
analyze our data statistically and find that EROs, as a class, 
have a significant submillimeter flux and produce much of the 
850~$\mu$m EBL.  In \S~\ref{secfaint} we use our NIR 
photometry and lensing corrections 
to calculate the ERO surface densities to $K'=24$. We 
analyze where in $K'$ magnitude the 850~$\mu$m flux
arises. Using the $I$, $J$, $K'$ colors, we differentiate
between evolved galaxies and dust-reddened starbursts to show
that the 850~$\mu$m light primarily arises in the latter class of 
object.

\section{Data and Sample Selection}
\label{secdata}

We constructed $K'$ catalogs for our three cluster fields
using data obtained with the 
Cooled Infrared Spectrograph and Camera for OHS
(CISCO; \markcite{moto98}Motohara et al.\ 1998) on the 
Subaru\footnote{The Subaru telescope is operated by the National 
Astronomical Observatory of Japan.} 8.2~m telescope. 
The detector is a $1024\times 1024$ HgCdTe HAWAII array with
a $0.111''$ pixel scale and an $\sim 2'\times 2'$ field-of-view.
The $K'$ exposures are 15.4~ks (A370), 3.84~ks (A851), and
15.4~ks (A2390).
$J$-band data were also obtained with the same instrument.
Details of the observations and data reduction can be found in
\markcite{cowie01}Cowie et al.\ (2001; A370 and A2390) and 
L.~L.~Cowie et al., in preparation (A851). 
The image quality of the $K'$ images is $0.66''$ FWHM. 
We used $K'$ magnitudes measured in $2''$ diameter apertures
to determine colors, and we used isophotal magnitudes to 
2\% of the peak surface brightness for total magnitudes.
We estimated the noise in the images by measuring aperture
magnitudes at a number of blank sky positions. The $5\sigma$ 
limit for the shallowest of the fields is $K'=21.25$;
we adopt this limit for the entire $K'$ sample selection.

We measured $2''$ diameter aperture $I$-band 
magnitudes for the $K'$ samples
from images obtained with either the Low-Resolution Imaging 
Spectrometer (LRIS; \markcite{oke95}Oke et al.\ 1995) or the Echellette 
Spectrograph and Imager (ESI; \markcite{epps98}Epps \& Miller 1998)
on the Keck\footnote{The W.~M.~Keck Observatory
is operated as a scientific partnership among the California
Institute of Technology, the University of California, and NASA,
and was made possible by the generous financial support of the
W.~M.~Keck Foundation.} 10~m telescopes. The data are described 
in \markcite{cowie01}Cowie et al.\ (2001; A370 and A2390) and
\markcite{smail99}Smail et al.\ (1999; A851).
The $2\sigma$ limit for the shallowest of the fields is $I=26$;
we adopt this limit for the entire sample. 

Finally, we measured 850~$\mu$m fluxes at the $K'$ source positions
from ultradeep SCUBA jiggle maps.
We used beam-weighted extraction routines that include both 
the positive and negative portions of the beam profile.
The total exposure times are 133.1~ks (A370), 59.3~ks (A851), and
78.1~ks (A2390). The $1\sigma$ sensitivities, including both fainter
sources and correlated noise, are 0.45~mJy (A370), 0.87~mJy (A851),
and 0.71~mJy (A2390). The data are described in 
\markcite{cowie02}Cowie, Barger, \& Kneib (2002).

\section{Submillimeter Properties of EROs and their Contribution
to the Submillimeter EBL}
\label{secebl}

%
%
\begin{inlinefigure}
\psfig{figure=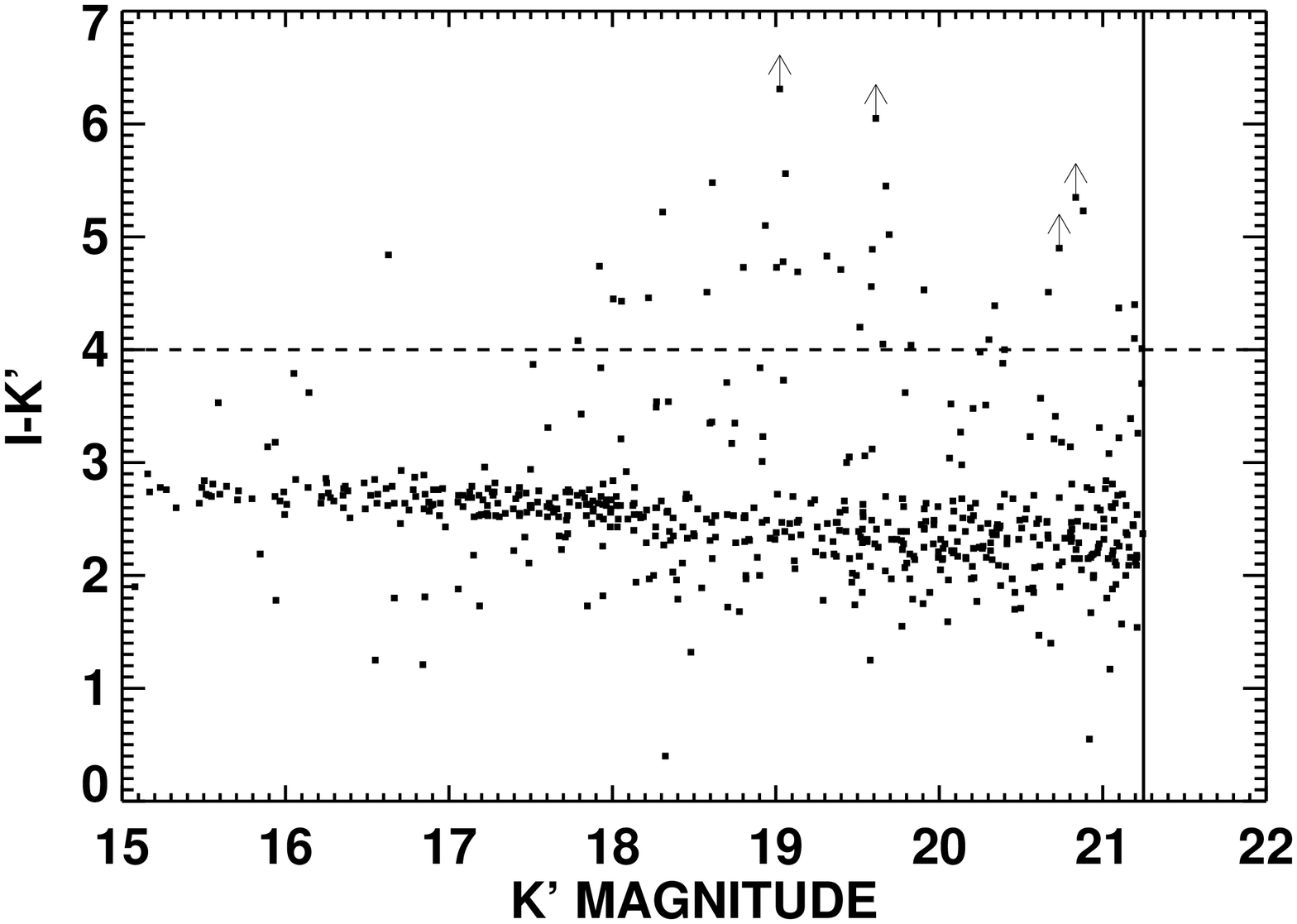,angle=90,width=3.5in}
\vspace{6pt}
\figurenum{1}
\caption{
$I-K'$ versus $K'$ for the combined $K'$ samples
in the A370, A851, and A2390 fields. The vertical line denotes
the $K'=21.25$ ($5\sigma$) limit adopted. The upward pointing arrows
denote sources that were not detected above $I=26$ ($2\sigma$).
}
\label{fig1}
\addtolength{\baselineskip}{10pt}
\end{inlinefigure}

Figure~\ref{fig1} shows the $I-K'$ versus $K'$ color-magnitude diagram.
Most of the sources are cluster members.
The cluster color-magnitude relationship is clearly visible
and matches the measured relations for clusters
at these redshifts (e.g., \markcite{kodama98}Kodama et al.\ 1998).
Thirty-eight galaxies have $I-K'>4$, classifying them as
EROs, and fifty-five have $I-K'>3.5$. We classify
galaxies with $3.5<I-K'<4$ as Very Red Objects (VROs).
The 11.5~arcmin$^2$ area covered by the $K'$ data is 
somewhat smaller than the 18.9~arcmin$^2$ area covered by the
submillimeter data. A disproportionate number of
the EROs and VROs lie in the A2390 cluster field (26 of the 38 sources
with $I-K'>4$, and 32 of the 55 sources with $I-K'>3.5$).
The excess may be associated with a very high-redshift,
evolved cluster lying behind A2390, as we discuss in \S~\ref{secfaint}. 

%
%
\begin{inlinefigure}
\psfig{figure=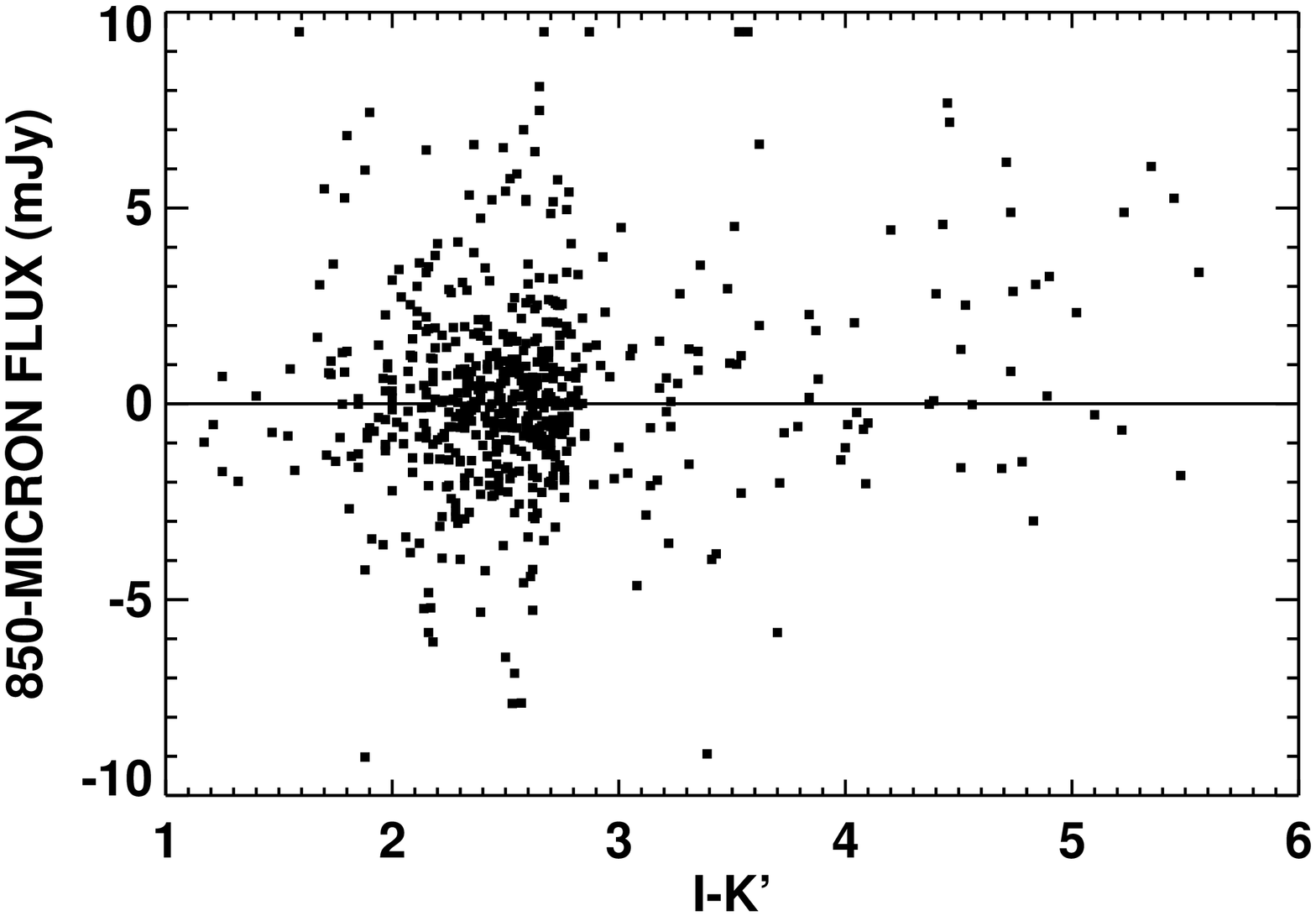,angle=90,width=3.5in}
\psfig{figure=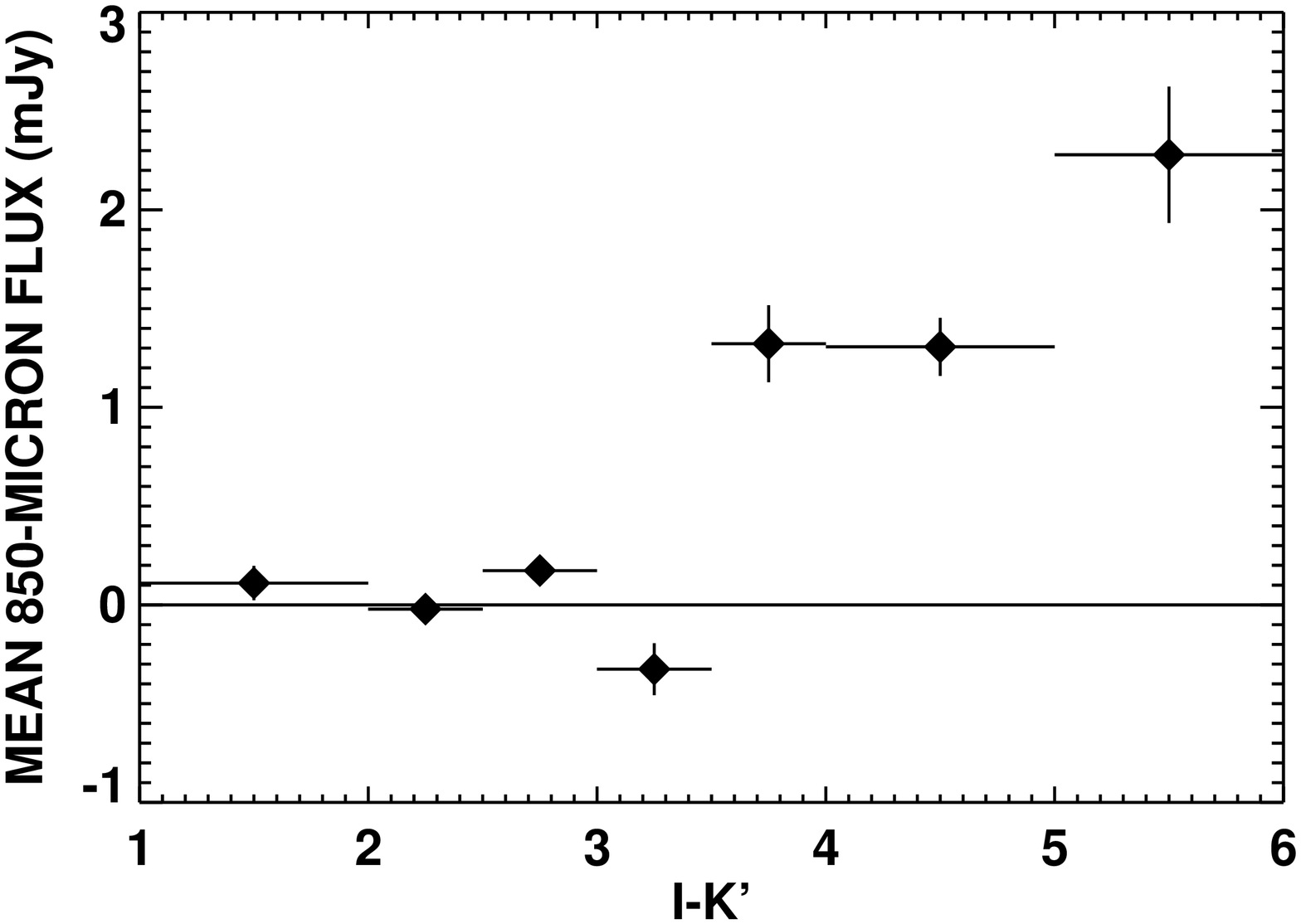,angle=90,width=3.5in}
\vspace{6pt}
\figurenum{2}
\caption{
(a) 850~$\mu$m flux and (b) error-weighted
mean 850~$\mu$m flux versus $I-K'$ color for the combined
$K'$ samples in the A370, A851, and A2390 fields.
In (a) any sources with 850~$\mu$m flux greater than 9.5 mJy are
plotted at 9.5 mJy. In (b) the vertical bars show the $1\sigma$
uncertainties and the horizontal bars show the color range
sampled in each bin.
}
\label{fig2}
\addtolength{\baselineskip}{10pt}
\end{inlinefigure}

The $15''$ SCUBA beam size means that we cannot be certain which 
$K'$-selected galaxy is the true counterpart to a submillimeter 
source. However, by determining the error-weighted mean 850~$\mu$m
flux of the $K'$ sources, we can
make statistically significant statements regarding the
relationship between submillimeter flux and source color. This
is analogous to what \markcite{peacock00}Peacock et al.\ (2000) 
did for optical starbursts. 

In Fig.~\ref{fig2}a we plot submillimeter flux versus $I-K'$ color,
and in Fig.~\ref{fig2}b we plot error-weighted mean 850~$\mu$m flux 
per color bin. Because submillimeter sources
are not removed from the jiggle maps, 
the complex beam profile produced by the nod and chop 
observing pattern (there are twice as many negative as positive 
response points, but at half the sensitivity) means that the
submillimeter flux distribution is non-Gaussian with an extended 
negative tail. A negative submillimeter flux will be measured for
a $K'$-selected source positioned in the negative portion of the 
beam profile corresponding to a submillimeter source. 
Thus, there are more high negative flux measurements
in Fig.~\ref{fig2}a than would be expected based on a simple
Gaussian model with the quoted errors. The average signal
of a set of random sources should, however, be zero, and the
significance of the result can be determined from sets of Monte Carlo 
simulations of the same number of points randomly distributed 
in the fields (more details can be found in 
\markcite{cowie02}Cowie et al.\ 2002).

At low values of $I-K'$ the error-weighted mean 850~$\mu$m fluxes
are consistent with zero; however, as $I-K'$ increases, there 
is a corresponding increase in the means (see Fig.~\ref{fig2}b).
The $I-K'>4$ galaxies have an error-weighted mean
of $1.58\pm 0.13$ mJy. The formal
error is based on the assumption of Gaussian statistics and
may underestimate the true error, but the result is highly
significant: 10000 Monte Carlo simulations yielded only
one case where 
the same number of randomly positioned objects had a value
equal or above the observed value.
We tested whether the signal is arising
from a small number of objects by eliminating the 
positive and negative tails: if the two (five) most
positive and two (five) most negative sources are removed from the
$I-K'>4$ sample, the signal becomes $1.40\pm 0.14$~mJy
($1.48\pm 0.15$~mJy). 
The $I-K'>4$ objects in the A2390 field 
have a value of $1.41\pm 0.18$~mJy,
while those in the A370 and A851 fields
have a value of $1.80\pm 0.20$~mJy,
so the mean submillimeter flux per ERO is consistent, even if we
break down the sample.
The $3.5<I-K'<4$ galaxies have an
error-weighted mean of $1.32\pm 0.19$ mJy; here 10000 simulations 
yielded only 85 cases with values at or above this level.

Since the sky surface brightness is conserved by the lensing 
process, we find the ERO population contributes 
$1.88\pm 0.16\times 10^4$~mJy~deg$^{-2}$ to the 850~$\mu$m EBL.
This value is $61$ percent of the total 850~$\mu$m EBL 
if the \markcite{puget96}Puget et al.\ (1996) value of 
$3.1\times 10^4$~mJy~deg$^{-2}$ is assumed, or $43$ percent 
if the \markcite{fixsen98}Fixsen et al.\ (1998) value
of $4.4\times 10^4$~mJy~deg$^{-2}$ is assumed. When the galaxies
with $3.5<I-K'<4$ are included, the contribution increases to
$2.59\pm 0.19 \times 10^4$~mJy~deg$^{-2}$ (83 or 59 percent of 
the total, respectively). 
Thus, the majority of the 850~$\mu$m EBL arises from 
the relatively small number of sources with $I-K'>3.5$. 

However, there may be systematic errors arising from clustering.
For the individual fields, the EBL contributions from the 
$I-K'>3.5$ sources are $0.81\pm 0.14\times 10^4$~mJy~deg$^{-2}$
(A370), $6.82\pm 0.42\times 10^4$~mJy~deg$^{-2}$ (A851), and 
$3.49\pm 0.55\times 10^4$~mJy~deg$^{-2}$ (A2390). 
Although A2390 has the largest number of $I-K'>3.5$ objects
(see \S~\ref{secfaint}), A851 has
a higher mean submillimeter signal per object.
If we exclude A2390 because of the anomalously high number
of red objects in this field, then
the $I-K'>3.5$ population contributes
$1.94\pm 0.15\times 10^4$~mJy~deg$^{-2}$ to the 850~$\mu$m EBL.
While the $I-K'>3.5$ populations in all three cluster fields are
significantly detected, it is clear that a large number
of fields would have to be averaged to produce a high precision
estimate of the EBL contribution, and the systematic errors 
in the present estimate may be as high as a factor of 2.

\section{The Breakdown of the ERO Contributions}
\label{secfaint}

We now analyze the ERO number counts to determine how 
representative our fields are and 
to estimate the ERO surface densities to fainter magnitudes than 
has previously been possible. The methodology used is described in 
more detail in \markcite{smith02}Smith et al.\ (2002),
who analyzed 10 cluster fields observed to shallower $K$ limits.

We used the LENSTOOL software package to determine the amplifications
of the EROs, assuming they were located at $z=2$. 
(The amplification varies only weakly with redshift for
any source beyond $z=1$ that has a modest amplification.)
LENSTOOL uses multiple-component mass distributions that describe
the extended potential wells of the clusters and their more
massive individual member galaxies 
(\markcite{kneib96}Kneib et al.\ 1996).
Details for the A370, A851, and A2390 clusters can be found in 
\markcite{kneib93}Kneib et al.\ (1993),
\markcite{seitz96}Seitz et al.\ (1996), and 
J.-P.~Kneib et al., in preparation, respectively. 

We also used LENSTOOL to determine the areas in the source plane
over which an object with a given source plane magnitude would
be seen. We created grids the 
size of our cluster field 
$K'$ images with $1''\times 1''$ grid elements and treated the grid 
points as objects in the source plane at
$z=2$. We brought the grids through the cluster 
mass distributions to determine the magnifications at each point.
Since each point represents one square arcsecond, the associated 
source plane area for a given magnitude is the sum of 
the points where an object of that magnitude could be detected.

To compute the ERO surface densities, we summed the reciprocals 
of the associated areas for all the sources in each magnitude bin. 
(Multiply imaged sources are correctly handled
since the source plane area is proportionally increased relative 
to the number of images.) The results 
are shown in Fig.~\ref{fig3}, where surface 
density per square degree per magnitude is plotted versus 
lensing-corrected (i.e., source plane) isophotal $K'$ magnitude.
The fact that Fig.~\ref{fig3} flattens at faint magnitudes 
indicates that by $K'=24$ we have already seen most of the $K'$ 
light density in the ERO population.

%
%
\begin{inlinefigure}
\psfig{figure=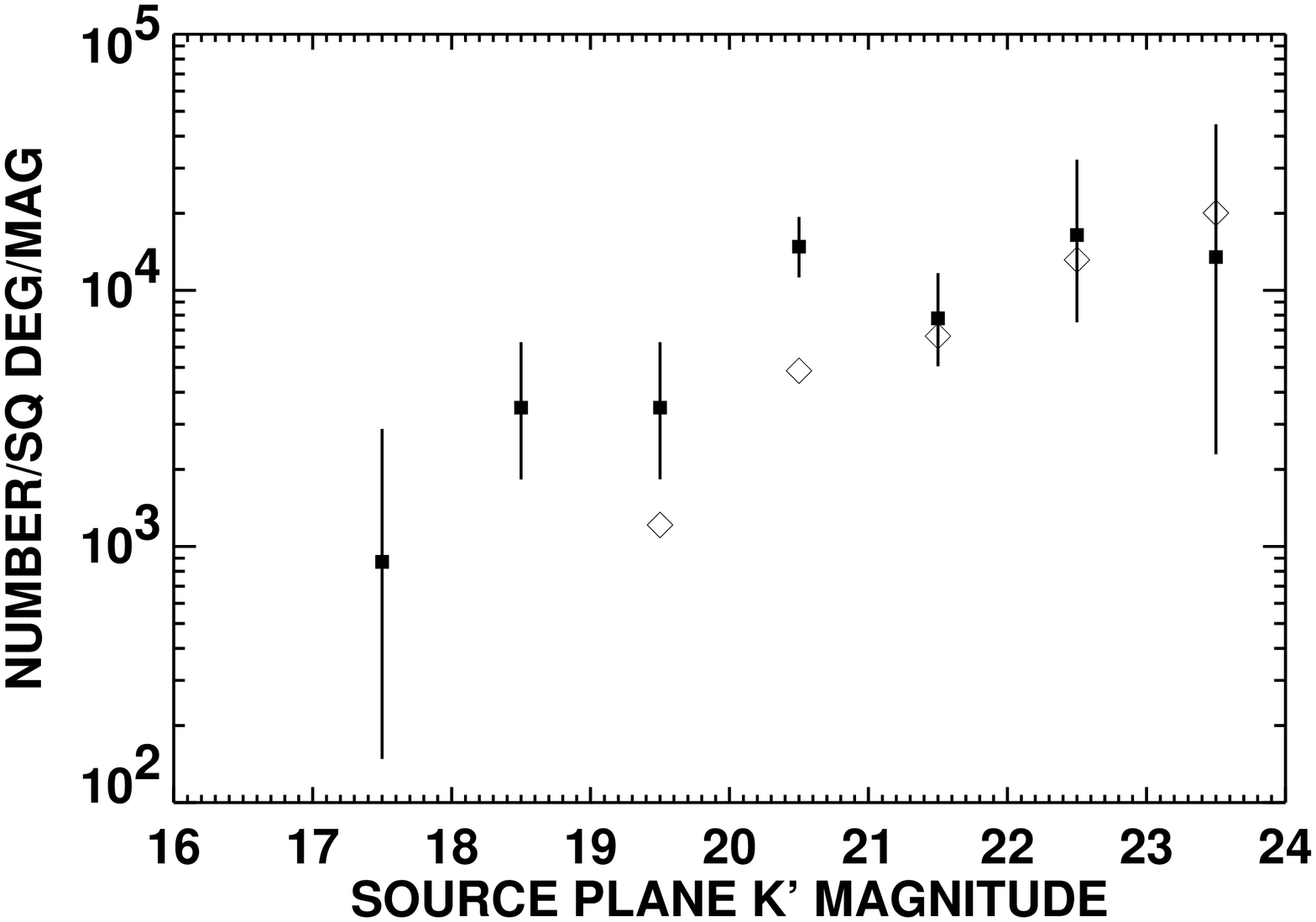,angle=90,width=3.5in}
\vspace{6pt}
\figurenum{3}
\caption{
Surface density per square degree per magnitude
versus lensing-corrected isophotal $K'$ magnitude for
the combined $K'$ samples in the A370, A851, and A2390 fields
with $I-K'>4$ (filled squares). The uncertainties are $1\sigma$,
based on a Poissonian distribution. The open diamonds show the
surface densities when the A2390 field is excluded.
}
\label{fig3}
\addtolength{\baselineskip}{10pt}
\end{inlinefigure}

At brighter $K'$ magnitudes our surface densities are higher 
than those determined by \markcite{smith02}Smith et al.\ (2002) and
by previous blank field surveys
(e.g., \markcite{barger99}Barger et al.\ 1999; 
\markcite{daddi00}Daddi et al.\ 2000). To $K'=21.6$ we 
find a cumulative surface density of 8.0 (6.6, 9.4) EROs
per square arcminute, where the parentheses show the $1\sigma$ 
limits. This is slightly more than three times the Smith et al.\ 
value of $2.5\pm 0.4$ arcmin$^{-2}$.
(Our $I-K'=4$ criterion is very similar to their $R-K>5.3$
criterion.) The reason for the excess lies in the large population 
of EROs in the A2390 field, where the
surface density is 19.7 (15.6, 23.8) arcmin$^{-2}$
to $K'=21.6$. If we restrict our analysis 
to the A370 and A851 cluster fields (the open diamonds in 
Fig.~\ref{fig3}), the surface density drops to 
3.4 (2.3, 4.8) arcmin$^{-2}$ to $K'=21.6$, which is 
comparable to the Smith et al.\ value.

Inspection of the $K'$ image of A2390
suggests that the ERO excess in this field corresponds
to a cluster of galaxies centered on a bright red galaxy
in the northwest corner of the image; this galaxy is
an X-ray point source (\markcite{fabian00}Fabian et al.\ 2000)
with a spectroscopic redshift of $z=1.467$ 
(\markcite{cowie01}Cowie et al.\ 2001).
A second X-ray source in the southwest corner has a similar 
probable spectroscopic redshift (Cowie et al.). 
If we assume that the excess of EROs is associated
with this cluster, then the cluster contains roughly 20 EROs to 
$K'=21.6$, making it an extremely large concentration of evolved 
galaxies at this redshift.

The error-weighted mean 850~$\mu$m flux per source decreases 
(though slower than linearly) with increasing source plane $K'$ 
magnitude: at mean $K'$ magnitudes of (16.9, 18.8, 20.6, 22.4)
we find error-weighted mean 850~$\mu$m fluxes of
($5.18\pm0.38, 1.55\pm0.29, 1.22\pm0.14, 0.47\pm0.20$)~mJy.
Although submillimeter flux depends only weakly on redshift, such
that higher redshift sources fade in $K'$ but not in 850~$\mu$m flux, 
sources that are fainter in $K'$ luminosity 
may be intrinsically fainter in 850~$\mu$m luminosity.
The dependence of 850~$\mu$m flux on 
ERO $K'$ magnitude means that brighter submillimeter surveys
are more likely to identify ERO counterparts.

The observed decrease in 850~$\mu$m flux with increasing $K'$ magnitude,
combined with the flatness of the ERO counts at the fainter 
magnitudes seen in Fig.~\ref{fig3}, implies that the 
largest contribution to the 850~$\mu$m EBL arises near the point at 
which the ERO number counts begin to flatten---around $K'=20$.
In the present data half of the 850~$\mu$m 
light arises in sources
with source plane magnitudes $K'<20$, and half arises in fainter 
sources, placing the median 850~$\mu$m source at 
$K'=20$. If A2390, with its larger number of bright EROs, is excluded, 
the split occurs at $K'=20.3$. When sources fainter than $K'=24$ 
are included, the split will occur at a fainter magnitude,
but the effect will be small.

We may also try to determine if the spectral energy distributions
(SEDs) of the sources correlate with the 850~$\mu$m
fluxes. If the 850~$\mu$m signal arises primarily from dust-reddened
starbursts rather than from old elliptical galaxies, then we would
expect objects with a smoother NIR SED to be the
primary source of the signal. (The red $I-K'$ colors of
old evolved galaxies are in large part due to the 4000~\AA\ break,
which produces a discontinuity in the NIR SEDs.)
As illustrated in Fig.~\ref{fig4}, we have used the
criterion $I-J=1.8(J-K')-1.2$ to separate redder $I-J$ galaxies
that follow the unevolved elliptical galaxy track
from bluer $I-J$ galaxies that lie in the dust-reddened starburst
region (see \markcite{pozzetti00}Pozzetti \& Mannucci 2000 for other
color selection criteria). Sources in the starburst
region could also be more recently formed galaxies or galaxies
with some residual star formation.

%
%
\begin{inlinefigure}
\psfig{figure=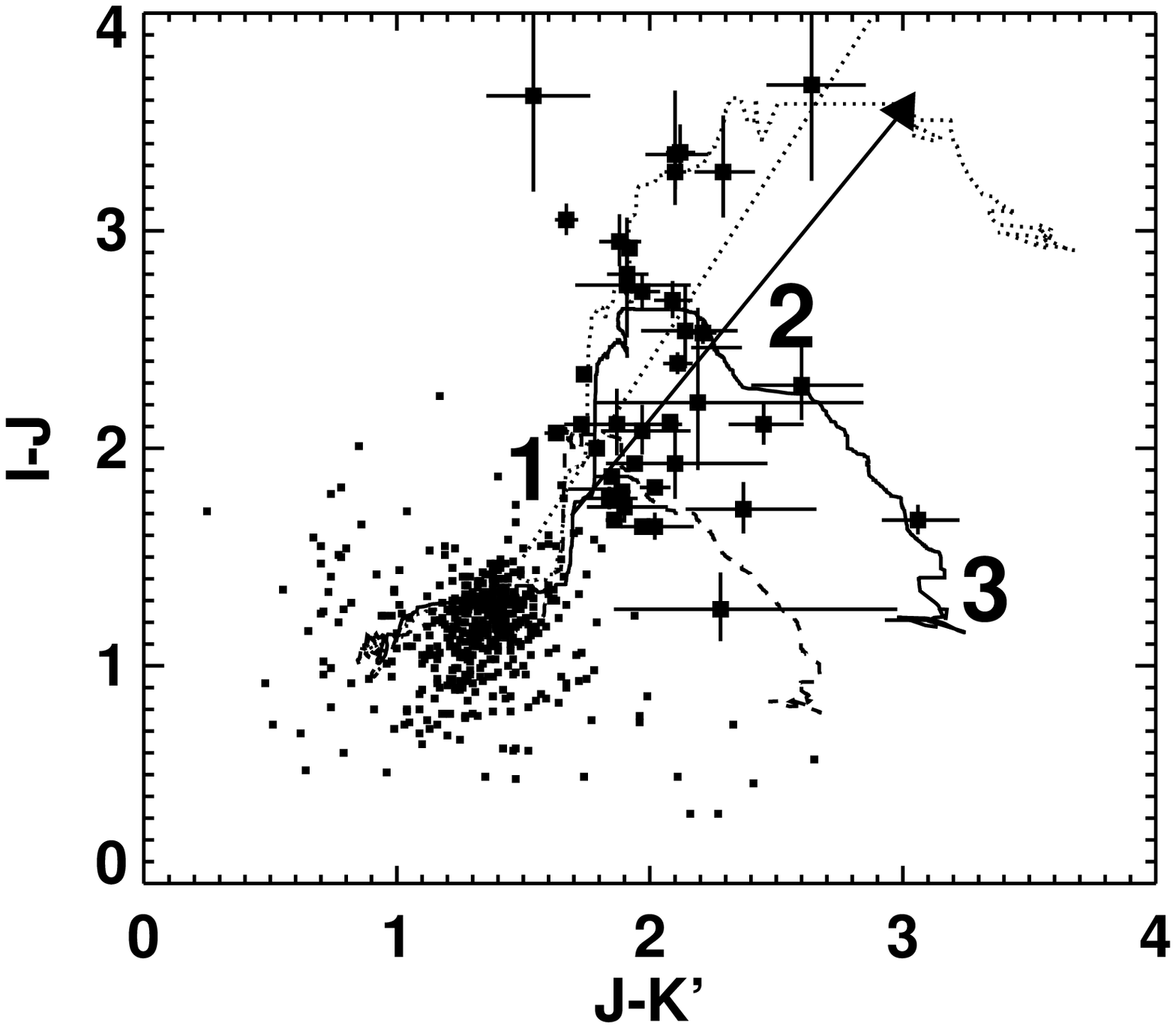,angle=90,width=4in}
\vspace{6pt}
\figurenum{4}
\caption{
$I-J$ versus $J-K'$ for the combined
$K'$ samples in the A370, A851, and
A2390 fields with measured $J$ magnitudes. Sources with $I-K'>3.5$
($I-K'<3.5$) are denoted by large (small) solid squares;
$1\sigma$ uncertainties are given for the $I-K'>3.5$ sources.
The dashed, solid, and dotted curves show, respectively, the tracks
for an unevolved Sb, Sa, and elliptical galaxy versus
redshift from $z=0$ to $z=3$. Redshifts $z=1$, 2, and 3 are marked
and labeled on the Sa track. The arrow shows the direction that
reddening would move a galaxy using a simple model where
extinction is inversely proportional to wavelength. The length of
the arrow corresponds to $A_V=10$. The dotted line $I-J=1.8(J-K')-1.2$
roughly separates objects that follow the unevolved elliptical galaxy
track (left of the dotted line) from objects located where
reddening could move a starburst galaxy (right of the dotted line).
}
\label{fig4}
\addtolength{\baselineskip}{10pt}
\end{inlinefigure}

Of the 55 galaxies with $I-K'>3.5$ in our sample,
39 have measured $J$ magnitudes. The error-weighted mean 850~$\mu$m
flux for all 39 objects is $0.90\pm0.12$~mJy. The 17
(44 percent) that satisfy the elliptical galaxy criterion
have $0.32\pm0.18$~mJy. In contrast, the
22 that lie on the dusty starburst side
have $1.38\pm0.16$~mJy. Thus, the bulk of the signal does appear 
to arise in the smaller number of galaxies that
could be dusty starbursts.

\section{Summary}

From an optical, NIR, and submillimeter survey of three massive 
lensing cluster fields we find 
that VROs and EROs mark, on average, strong submillimeter emitters. 
The $I-K'>3.5$ population contributes 
$1.94\pm 0.15\times 10^4$~mJy~deg$^{-2}$ to the 850~$\mu$m EBL
if we exclude the A2390 field because of the anomalously high number
of red objects, and $2.59\pm 0.19\times 10^4$~mJy~deg$^{-2}$
if we include the A2390 field. These contributions are, respectively, 
$44-62$ percent and $59-83$ percent of
the 850~$\mu$m EBL (the ranges depend on the measurement of the EBL
assumed) and are much larger than the $12-16$ percent
from galaxies containing bright AGN
(\markcite{barger01}Barger et al.\ 2001).
ERO clustering is an important effect that must be accounted 
for if we are to obtain a better estimate.
The median magnitude of the sources giving rise to the
EBL is $K'=20$ to $K'=20.3$, depending on the cluster sample 
used. The 850~$\mu$m EBL contribution primarily arises in the 
subset of the population corresponding to dusty starbursts.

\newpage

\acknowledgements
We thank an anonymous referee for helpful comments. 
AJB gratefully acknowledges support from NSF grant AST-0084847
and from the University of Wisconsin Research Committee with funds
granted by the Wisconsin Alumni Research Foundation.

\end{document}